\documentclass[12pt,a4paper,titlepage,oneside]{article}
\usepackage{amsmath}
\usepackage{graphicx}%

\begin{document}
\begin{center}
  \textbf{\Large {The modified electromagnetism and the emergent
      longitudinal wave}}
\end{center}

{\bf A. I. Arbab$^{1}$} and {\bf M. Al-Ajmi$^{2}$} \\

$^1$Department of Physics, Faculty of Science, University of
  Khartoum, P.O. Box 321, Khartoum 11115, Sudan \\
$^2$Department of Physics, College of Science, Sultan Qaboos
  University , P.O. Box 36, Alkhod 123, Oman \\

{\bf email:} aiarbab@uofk.edu, mudhahir@squ.edu.om \\


\section*{Abstract}
The classical theory of electromagnetism has been revisited and the
possibility of longitudinal photon wave is explored. It is shown that
the emergence of longitudinal wave is a consequence of Lorenz gauge
(condition) violation. Proca, Vlaenderen \& Waser and Arbab theories
are investigated. The different approaches are compared to each other
and the relevant equations are combined. The telegrapher's equation
can be obtained and with a specific choice of the function a Klein-Gordon
equation of massive scalar field can be obtained. When the Lorentz gauge is
violated by introducing the first order time derivative the emergence of the
photon mass and the relevant longitudinal wave for the electromagnetic wave is apparnt. \\



\baselineskip=20pt
\section{Introdution}
Maxwell's equations are one of the biggest edifices that mankind had
come to establish. Maxwell unified electromagnetism with light in a
set of consistent equation dubbed in Maxwell equation. These equation
prematurely predicted that speed of light in vacuum is constant. In an
elegant theory, Salam, Weinberg and Glashow unified electromagnetism
with weak interaction \cite{salam1, salam2, salam3}. The ultimate goal
of unifying all interactions into a single universal theory is
underway. Some theoretical prejudices advocated by Proca in extending
Maxwell's theory to encompass massive photons was expounded in 1967
\cite{proca}.  This later theory will be of importance, at cosmic
scale, as some compelling investigation usher in that direction. Of
these investigation is the missing mass (dark matter) in the universe.
It is remarked by Schrodinger that if the photon had a mass then the
Black-Body formula will be modified and the energy density of these
photons would be 3/2 times the ordinary one \cite{schrod}. This is
because these photon would have a longitudinal polarization besides
the transverse one. Longitudinal waves have been shown to have high
penetration power. This character rendered them to be of great
importance and will find wider applications in communication systems.
Some endeavors to generalize Maxwell's equations to remedy these
shortcoming have been proposed by following a quaternionic
generalizations. In the framework of these attempts, the shortcoming
present in the original Maxwell's formulation can be avoided
\cite{satti}. In this work, we will outline the different approaches
and explore these interrelations. We will see that whenever the Lorenz
gauge (condition) is violated, longitudinal waves will be inevitable.
We also analyze the status of the experimental evidence supporting the
presence of these waves.
\section{Maxwell's equation}
Maxwell's equations describe the behavior of the electric and magnetic
fields. These equations are \cite{griffith}
\begin{equation}
\vec{\nabla}\cdot\vec{E} =\frac{\rho}{\varepsilon_0}\,,\qquad   \vec{\nabla}\cdot\vec{B}=0\,,
\end{equation}
and
\begin{equation}
\vec{\nabla}\times\vec{E}=-\frac{\partial\vec{B}}{\partial t}\,,\qquad \vec{\nabla}\times\vec{B} =\mu_0\vec{J}+\frac{1}{c^2}\frac{\partial\vec{E}}{\partial t}\,.
\end{equation}

\section{Vector and scalar potential}
The electric and magnetic fields are defined as
\begin{equation}
\vec{E}=-\vec{\nabla}\varphi-\frac{\partial\vec{A}}{\partial t}\,,\qquad\qquad \vec{B}=\vec{\nabla}\times\vec{A}\,.
\end{equation}
It is believed that vector and scalar potentials, $\vec{A}$ and
$\varphi$, are more fundamental than $\vec{E}$ and $\vec{B}$. For this
reason, we describe the electromagnetic interaction in terms of these
potentials, rather than the electric and magnetic fields. This is
apparent, at the quantum level, in the Aharonov-Bohm effect where when
two beams of electrons are sent through region with same $\vec{B}$,
but different $\vec{A}$, experience an interference pattern when
collected at the detector \cite{ahar}. This occurs because the phase
factor of the electron wavefunction depends on $\vec{A}$.

Substituting eq.(3) in (1) and (2), we obtain \cite{griffith}
\begin{equation}
\frac{1}{c^2}\frac{\partial^2\varphi}{\partial t^2}-\nabla^2\varphi-\frac{\partial}{\partial t}\left(\vec{\nabla}\cdot\vec{A}+\frac{1}{c^2}\frac{\partial\varphi}{\partial t}\right)=\frac{\rho}{\varepsilon_0}\,,
\end{equation}
and
\begin{equation}
\frac{1}{c^2}\frac{\partial^2\vec{A}}{\partial t^2}-\nabla^2\vec{A}+\vec{\nabla}\left(\vec{\nabla}\cdot\vec{A}+\frac{1}{c^2}\frac{\partial\varphi}{\partial t}\right)=\mu_0\vec{J}\,.
\end{equation}
These are the Maxwell's equations in terms of the vector and scalar potentials.
\subsection{Lorenz gauge}
The electric and magnetic fields are invariant under the gauge
transformations,
\begin{equation}
\vec{A}\,'=\vec{A}+\vec{\nabla}\Lambda\,\,,\qquad\qquad  \varphi\,'=\varphi-\frac{\partial\Lambda}{\partial t} \,.
\end{equation}
This allows the freedom of taking the (Lorenz gauge) condition,
\begin{equation}
\vec{\nabla}\cdot\vec{A}+\frac{1}{c^2}\frac{\partial\varphi}{\partial t}=0\,.
\end{equation}
In this case one has the condition that
\begin{equation}
\frac{1}{c^2}\frac{\partial\Lambda}{\partial t}-\nabla^2\Lambda=0\,.
\end{equation}
It is interesting to see that the electric and magnetic fields can not
be uniquely specified by $\vec{A}$ and $\varphi$. For this reason
these two fields are treated as mere mathematical constructs and bear
no physical meaning, but $\vec{E}$ and $\vec{B}$ do.
\section{Vlaenderen and Waser approach}
Vlaenderen and Waser have recently adopted a biquaternionic
formulation of Maxwell's equation and introduced a scalar function
($S$) that is a measure of the violation of Lorenz gauge \cite{waser}.
They remarked that this scalar field mimics the Maxwell's displacement
current. For this case, they assumed that
\begin{equation}
\vec{\nabla}\cdot\vec{A}+\frac{1}{c^2}\frac{\partial\varphi}{\partial t}=S\,.
\end{equation}
Accordingly, they obtain the following equations

\begin{equation}
\vec{\nabla}\cdot\vec{E} =\frac{\rho}{\varepsilon_0}-\frac{\partial S}{\partial t}\,,\qquad\vec{\nabla}\cdot\vec{B}=0 \,.
\end{equation}
and
\begin{equation}
\vec{\nabla}\times\vec{E}=-\frac{\partial\vec{B}}{\partial t}\,,\qquad \vec{\nabla}\times\vec{B} =\mu_0\vec{J}+\frac{1}{c^2}\frac{\partial\vec{E}}{\partial t}+\vec{\nabla} S.
\end{equation}
The scalar $S$ satisfies the wave equation
\begin{equation}
\frac{1}{c^2}\frac{\partial^2 S}{\partial t^2}-\nabla^2S=\mu_0\left(\frac{\partial\rho}{\partial t}+\vec{\nabla}\cdot\vec{J}\right)=0\,,
\end{equation}
upon using the continuity equation.  Besides this, $S$ is found to
represent an electroscalar wave having a Poynting vector $S\vec{E}$,
i.e., along the electric field direction.
\section{Proca's equations}
Proca generalized Maxwell's equations by introducing a massive photon filed. The resulting equations are \cite{proca}
\begin{equation}
\vec{\nabla}\cdot\vec{E} =\frac{\rho}{\varepsilon_0}-\mu_\gamma^2\varphi\,,\qquad\qquad\vec{\nabla}\cdot\vec{B}=0 \,,
\end{equation}
and
\begin{equation}
\vec{\nabla}\times\vec{E}=-\frac{\partial\vec{B}}{\partial t}\,,\qquad \vec{\nabla}\times\vec{B} =\mu_0\vec{J}+\frac{1}{c^2}\frac{\partial\vec{E}}{\partial t}-\mu_\gamma^2\,\vec{A}\,.
\end{equation}
\section{Universal quantum equations}
We have recently introduced a system of unified quantum wave equations
describing matter waves. Like Maxwell's formulation, in this
formulation the particle is described by scalar and vector potentials.
These equations read \cite{uqe}
\begin{equation}
\frac{1}{c^2}\frac{\partial^2\psi}{\partial t^2}-\nabla^2\psi +\frac{2m}{\hbar}\frac{\partial\psi}{\partial t}+\frac{m^2c^2}{\hbar^2}\,\psi=0\,,
\end{equation}
and
\begin{equation}
\frac{1}{c^2}\frac{\partial^2\vec{\psi}}{\partial t^2}-\nabla^2\vec{\psi} +\frac{2m}{\hbar}\frac{\partial\vec{\psi}}{\partial t}+\frac{m^2c^2}{\hbar^2}\,\vec{\psi}=0\,.
\end{equation}
Here $\psi$ and $\vec{\psi}$ are the scalar and vector fields
(potentials) representing the material wave nature endorsed by de
Broglie of a given particle (mass). We assume that any wave that
carries a material nature should satisfy these equations. When $m=0$
the material nature disappears and we obtain a wave traveling at the
speed of light. \\
The Lorentz gauge  keeps equations Lorentz invariant under Lorentz transformation.
Equations (15, 16) are Lorentz breaking equations because of the first order
time derivative implying that the Lorentz footing of coordinates is broken.
However, these equations are invariant under other class of transformation
as will be demonstrated in the next section.
\section{The analogy among these paradigms}
Comparing eqs.(10) \& (11) with eqs.(13) \& (14) yields \footnote{We consider here $-S$ instead of $S$}
\begin{equation}
\mu^2_\gamma\varphi=-\frac{\partial S}{\partial t}\,,\qquad \qquad\mu_\gamma^2\vec{A}=\vec{\nabla}S\,.
\end{equation}
Differentiation partially the first equation w.r.t. $t$, taking the divergence of the second equation and subtracting the resulting two equations yield
\begin{equation}
\frac{1}{c^2}\frac{\partial^2 S}{\partial t^2}-\nabla^2S+\mu^2_\gamma S=0\,.
\end{equation}
This is the familiar Klein-Gordon equation of massive spin zero
particle. Thus, if the photon had a non-zero mass, as hypothesized by
Proca, then this photon wave must be governed by eq.(18).

 Integrating the second equation in  eq.(17) yields
\begin{equation}
S=\int\mu^2_\gamma\vec{A}\cdot\vec{dr}\,.
\end{equation}
But from Stokes theorem,$\int\vec{A}\cdot\vec{dr}=\int \vec{\nabla}\times\vec{A}\cdot d\vec{S}=\phi_m\,$, where $\phi_m$ is the magnetic flux, one has
\begin{equation}
S=\mu^2_\gamma\phi_m+f(t)\,.
\end{equation}
Therefore, eq.(17) can be written as
\begin{equation}
\varphi=-\frac{\partial \phi_m}{\partial t}-\frac{\partial f}{\partial t}\,,
\end{equation}
we consider here $f(t)=\rm constant$.  The above equation is nothing
but the \emph{electromotive force}. Hence, the scalar potential
$\varphi$ represents the electromotive force. It is evident from
eq.(20) that if the Lorenz gauge is satisfied, then the photon must be
massless ($\mu^2_\gamma=0$), or otherwise $\phi_m=0$\,.  Moreover, the
non-satisfaction of Lorenz gauge would lead to massive photon and
non-zero magnetic flux. Equation (18) and (20) imply that the magnetic
flux satisfies the Klein-Gordon equation, where the flux has a mass,
$\mu^2_\gamma\,$. Thus, the magnetic flux would propagate as a bosonic
particle with spin zero.
If we assume now the scalar and vector potentials satisfy the matter
wave equation, i.e., eq.(15), then
\begin{equation}
\vec{\nabla}\cdot\vec{A}+\frac{1}{c^2}\frac{\partial\varphi}{\partial t}=-\frac{2m}{\hbar}\varphi\,.
\end{equation}
This equation can be compared with eq.(9), in which case,
$S=-\frac{2m}{\hbar}\,\varphi$. If we compare eqs.(9), (17), (18) and
(22) we will get
\begin{equation}
\frac{1}{c^2}\frac{\partial^2\varphi}{\partial t^2}-\nabla^2\varphi+\mu^2_\gamma\varphi=0\,.
\end{equation}
This implies that the scalar potential of the massive photon,
$\varphi$, satisfies the Klein-Gordon equation.

Now  differentiating partially eq.(22) with respect to time and using (3) yield
\begin{equation}
\frac{1}{c^2}\frac{\partial^2\varphi}{\partial t^2}-\nabla^2\varphi+\frac{2m}{\hbar}\frac{\partial\varphi}{\partial t}-\vec{\nabla}\cdot\vec{E}=0\,.
\end{equation}
Taking the gradient of eq.(22) and using eq.(3) and the second
equation in eq.(2), we obtain
\begin{equation}
\frac{1}{c^2}\frac{\partial^2\vec{A}}{\partial t^2}-\nabla^2\vec{A}+\frac{2m}{\hbar}\frac{\partial\vec{A}}{\partial t}+\frac{2m}{\hbar}\vec{E}-\mu_0\vec{J}=0\,,
\end{equation}
where we have used the vector identity, $\vec{\nabla}\times(\vec{\nabla}\times\vec{A})=\vec{\nabla}(\vec{\nabla}\cdot\vec{A})-\nabla\,^2\vec{A}$.

It is interesting to see that if the photon had a mass ($m\ne0$) in
vacuum, then the Lorenz gauge in eq.(22) would not be satisfied. This
encourages us to associate a non-zero mass of the photon to the
breaking of Lorenz gauge.

In vacuum, $\rho=0$ and $\vec{J}=0$, so that eqs.(24) and (25), using eq.(3),  become
\begin{equation}
\frac{1}{c^2}\frac{\partial^2\varphi}{\partial t^2}-\nabla^2\varphi+\frac{2m}{\hbar}\frac{\partial\varphi}{\partial t}=0\,.
\end{equation}
Taking the gradient  of eq.(22) and using eq.(3) and the second equation in eq.(2), we obtain
\begin{equation}
\frac{1}{c^2}\frac{\partial^2\vec{A}}{\partial t^2}-\nabla^2\vec{A}-\frac{2m}{\hbar}\vec{\nabla}\varphi=0\,.
\end{equation}
Therefore, eqs.(26) and (27) are the corresponding equations
describing massive photon in vacuum. Adding the time derivative of
eq.(27) to the gradient of eq.(26), using eq.(3), imply that both
$\vec{E}$ and $\vec{B}$ travel at speed of light, viz.,
\begin{equation}
\frac{1}{c^2}\frac{\partial^2\vec{E}}{\partial t^2}-\nabla^2\vec{E}=0\,\,,\qquad \frac{1}{c^2}\frac{\partial^2\vec{B}}{\partial t^2}-\nabla^2\vec{B}=0\,.
\end{equation}
This is a quite surprising result. It is evident from eqs.(26) and
(27) that spatial and temporal variations of the scalar field is the
source of the $\varphi$ and $\vec{A}$ waves.

Now apply the gauge transformation, eq.(6), in eq.(22) to get
\begin{equation}
\frac{1}{c^2}\frac{\partial^2\Lambda}{\partial t^2}-\nabla^2\Lambda+\frac{2m}{\hbar}\frac{\partial\Lambda}{\partial t}=0\,.
\end{equation}
This equation is know as the Telegrapher's equation. It describes the
electric signals in telephone and telegraph lines. However, by making
the substitution, $\Lambda (r, t)=e^{-(mc^2t/\hbar)}\chi(r, t)$,
eq.(29) can be written as
\begin{equation}
\frac{1}{c^2}\frac{\partial^2\chi}{\partial t^2}-\nabla^2\chi+\frac{m^2c^2}{\hbar^2}\,\chi=0\,.
\end{equation}
which is the ordinary Klein-Gordon equation of massive scalar field
$\chi$. This equation is a consequence of gauge invariance.
\subsection{The combined gauge transformation}
We will consider here the combined gauge transformations in which both
$\varphi$ \& $\vec{A}$, as well as, $\rho$ \& $\vec{J}$ are
transformed. Interestingly, Proca equations are invariant under these
combined gauge transformations. These are given by
\begin{equation}
\vec{A}\,'=\vec{A}+\vec{\nabla}\Lambda\,, \qquad \qquad\varphi\,'=\varphi-\frac{\partial\Lambda}{\partial t}\,,
\end{equation}
and
\begin{equation}
\vec{J}\,'=\vec{J}+\mu^2_\gamma\varepsilon_0c^2\,\vec{\nabla}\Lambda\,,\qquad \qquad \rho\,'=\rho-\mu^2_\gamma\varepsilon_0\,\frac{\partial\Lambda}{\partial t}\,.
\end{equation}
Under these gauge transformations, Proca's equations become gauge
invariant and that the mass term is no longer problematic. Moreover,
these transformations leave the system of the continuity equations
invariant as well, as long as $\Lambda$ satisfies the wave equation in
eq.(8) \cite{widat}.

\section{The status of experimental evidence of longitudinal wave and photon mass}
There are some appealing experiments that amount to the existence of
longitudinal wave. An electric longitudinal wave has been observed by
Miyaji \emph{et al.} while studying the transverse wave\cite{japan}.
This wave is characterized by the vanishing magnetic component.

The enormous success in quantum electrodynamics has lead to the
acceptance of the concept of massless photons. Also, the nonzero
photon mass would give rise to a wavelength dependence of the speed of
light in free space, a more rapid (exponential or Yukawa) falloff of
magnetic dipole fields with distance than the usual inverse cube
dependence \cite{[13]}. This could be attributed to the possibility of
existing longitudinal electromagnetic waves. This wave is
characterized by the vanishing magnetic and electric field components
\cite{[14]}. A massive photon is equivalent to a violation of
Coulomb's Law \cite{[15]}. Hence, a Yukawa factor $e^{-r/\lambda} $ in
the \textit{1/r} electrostatic and magnetostatic potential terms would
appear. The magnetic field term will contain correction terms related
to the de Broglie wavelength $\lambda _{\gamma } =\hbar /m_{\gamma}c.$
Another way to test the existence of the photon mass and hence the
longitudinal waves is the limit of the power in the inverse of the
distance distance in Coulomb law, namely $1/r^{2+\varepsilon }.$ One
of the consequences of $\varepsilon \ne 0$ is the decrease of the
electromagnetic waves as the wavelength increases which is due to the
smallness of $\varepsilon $. A limit of $\varepsilon =5.1\times
10^{-20} $ was set by Lou \textit{et al} \cite{[16]}. According to the
uncertainty principle the upper limit on the photon mass can be
obtained from $m_{\gamma} \approx \hbar /(\Delta t{\rm \; }c^{2} )$.
Using the age of the universe as 10${}^{10}$ years one obtains
$m_{\gamma} \approx 10^{-66} g$.

De Broglie noticed that photon mass leads to a larger speed of violet
light than that of the red one \cite{[17]}. He concluded that during
the eclipse in a double star system the color of the appearing star
would change from violet to red. On the other hand, Schrodinger
pointed out that magnetic field of the Earth would be exponentially
cut off at distances of the order of the photon Compton wave length
\cite{[18]}. From the observed altitude of auroras he concluded that
$\lambda _{\gamma } >10^{4} $. Other data have shown more limits also.

Because of limitations involved in experiments performed on the Earth,
astrophysical measurement could promise better limits on photon
masses. One of the consequences of photons being massive is the
dispersion of star light. According to measurements performed on
planet Jupiter the upper limit of Compton wavelength is $\lambda
_{\gamma } =3.14\times 10^{11} $cm \cite{[19]}.  Based on analyzing
the Earth's magnetic field the upper limit on photon mass is
$m_{\gamma} =1\times 10^{-48} $g \cite{[20]}. Also from geomagnetic
field Fischbach \emph{et al.} found a limit on the upper mass of the
photon as $m_{\gamma } \le 10^{-48} $g \cite{[21]}. According to Proca
equation, the velocity of light depends on its frequency. From the
measurement on the radio signals from the Crab Nebula it was found
that the upper limit on the mass of the photon is $m_{\gamma } \le
10^{-44} $g \cite{[22]}.  Also, hydrodynamic waves coupling to the
interplanetary magnetic fields could be useful in finding the mass of
the photon. Ryutov (1997) set a limit of $m_{\gamma } \le 10^{-51} $g
based on these studies \cite{[23]}. Other limits on the photon mass
are obtained from studying the current density and plasma of the
interstellar media ($m_{\gamma } \le 10^{-53} $g) \cite{[24]}.

Modern theories that appeal to the concept of spontaneous symmetry
breaking assume that particles, which are massless above a certain
critical temperature $T_{c} $, acquire mass below this temperature.
This can be applied to photons.  From the cryogenic photon mass
experiment performed by Ryan \textit{et al} an upper limit mass on
photons were found to be $m_{\gamma } =1.5\times 10^{-42} $. From the
symmetry breaking point of view and the experiment, one can speculate
that photons can be massless above a certain temperature \cite{[25]}.

Superconductivity is characterized by perfect conductivity and perfect
diamagnetism.  BCS theory can explain both of these fundamental
properties of superconductors \cite{[26]}. However, the London
equations can be easily derived and used to understand the expulsion
of flux within the material without appealing to the broader BCS
theory \cite{[27]}. Such a derivation relies solely on postulating
that the Cooper pairs in the superconducting state maintain their
ground state momentum of $p = 0$.  It has been possible to identify an
effective photon mass that is directly related to the London
penetration depth of the applied field \cite{[28]}.

In 2008 a Russian group performed a measurement on the volt value
triggered by the radiation coming from solar eclipse. It was noticed
the registered signal does not depend on the orientation of the
tourmaline plates used in front of the detector \cite{[29]}. This
could be an indication of the longitudinal wave component and, hence,
the non-zero mass of the photon.

Another experimental evidence for the existence of the massive photons
is the Tesla coil. The coil is a device which produces high frequency
alternating current with high voltage and low current. The coil is
used to produce energy which can be transmitted to a device without
any wires. A neon light can lit without wiring connection. This can be
explained in terms of the longitudinal component of the light wave.

\section{Conclusion}
In this paper, the possibility of longitudinal photon wave is explored
from the electromagnetic theory. We revisited Proca, Vlaenderen \& Waser and Arbab theories.
We studied the equations obtained from these approaches and combined them together.
The different approaches are compared to each other
and the relevant equations are combined. It is shown, then, that if
the Lorentz gauge invariance is violated the photon can gain mass and the longitudinal part of
the electromagnetic wave can appear.

\end{document}